\documentclass{article}

\usepackage[english]{babel}

\usepackage[letterpaper,top=2cm,bottom=2cm,left=3cm,right=3cm,marginparwidth=1.75cm]{geometry}

\usepackage{amsmath}
\usepackage{tabularx}
\usepackage{amssymb}
\usepackage{graphicx}
\usepackage{authblk}
\usepackage{array}
\usepackage{subfig}
\usepackage{booktabs}
\usepackage{pifont}
\usepackage{graphicx}
\usepackage{adjustbox}
\usepackage[colorlinks=true, allcolors=blue]{hyperref}

\title{Evaluating LLM-based Workflows for Switched-Mode Power Supply Design}

\author[1]{Simon Nau}
\author[2]{Jan Krummenauer}
\author[1,3]{Andr\'{e} Zimmermann}

\affil[1]{Robert Bosch GmbH, Cross-Domain Computing Solutions, Leonberg, Germany}
\affil[2]{University of Stuttgart, Institute for Micro Integration (IFM), Stuttgart, Germany}
\affil[3]{Hahn-Schickard, Stuttgart, Germany}
\affil[ ]{\texttt{Simon.Nau@de.bosch.com}}

\begin{document}
\date{}
\maketitle

\begin{abstract}
Large language models (LLMs) have great potential to enhance productivity in many disciplines, such as software engineering. However, it is unclear to what extent they can assist in the design process of electronic circuits. This paper focuses on the application of LLMs to switched-mode power supply (SMPS) design for printed circuit boards (PCBs).
We present multiple LLM-based workflows that combine reasoning, retrieval-augmented generation (RAG), and a custom toolkit that enables the LLM to interact with SPICE simulations to estimate the impact of circuit modifications. 
Two benchmark experiments are presented to analyze the performance of LLM-based assistants for different design tasks, including parameter tuning, topology adaption and optimization of SMPS circuits.
Experiment results show that SPICE simulation feedback and current LLM advancements, such as reasoning, significantly increase the solve rate on 269 manually created benchmark tasks from 15\% to 91\%. Furthermore, our analysis reveals that most parameter tuning design tasks can be solved, while limits remain for certain topology adaption tasks. Our experiments offer insights for improving current concepts, for example by adapting text-based circuit representations.\\
\end{abstract}

\section{Introduction}

\begin{figure}[h]
  \centering
  \includegraphics[width=\linewidth]{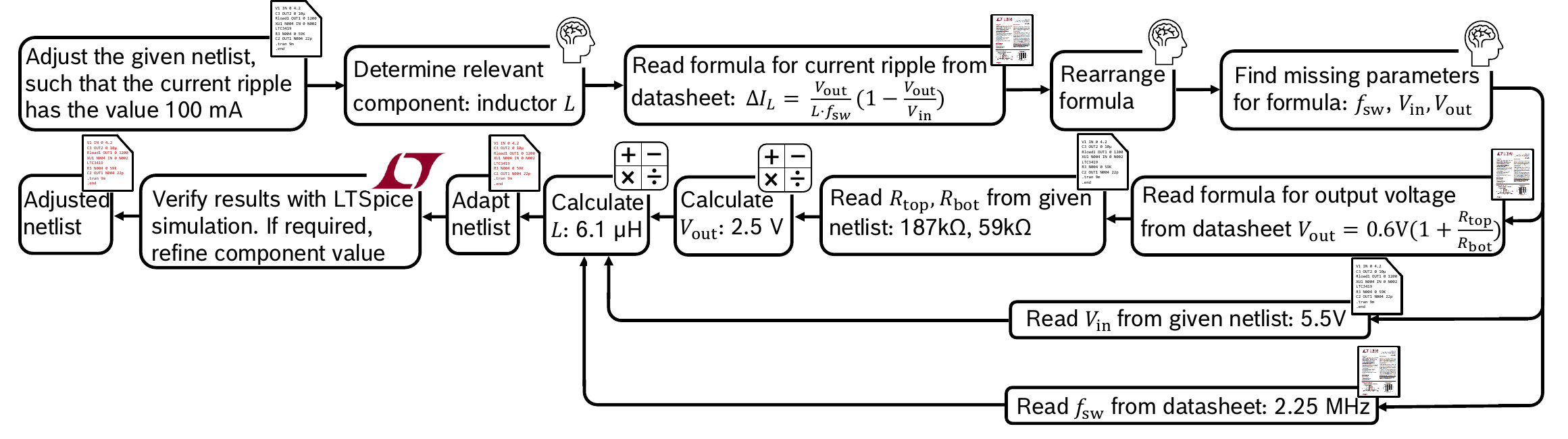}
  \caption{Required reasoning chain to solve the following exemplary SMPS design task: "Adjust the given netlist, such that the
current ripple has the value 100 mA".  The netlist and complete input prompt can be seen in Figure \ref{fig:colored_prompt}. The controller's datasheet and SPICE simulations are used as external information sources.\\
$\Delta I_{L}$: ripple current, $f_s$: switching frequency, $V_{\textrm{out}}$: output voltage, $V_{\textrm{in}}$: input voltage, $R_\textrm{top}$ and $R_\textrm{bot}$: resistors in feedback path}
\label{fig:thinking_steps}
\end{figure}

Large language models (LLMs) have achieved remarkable results in a broad field of applications. They revolutionize software development by automating significant parts of the development process, increasing speed and efficiency. Previous research aims to translate this success into the field of electronic design automation (EDA) \cite{chipchat, chipgpt, verigen, autochip, verilogeval, rtlfixer, rtllm, gpt4aigchip, rtlcoder, spicepilot, analogcoder, ladac, ampagent,  ledro, wiseeda, amsnet2, amsbench, masalachai}, with a main focus on the development of integrated circuits (ICs). Much of this work centers on digital ICs, by applying LLMs for hardware description language (HDL) code generation, particularly using Verilog \cite{chipchat, chipgpt, verigen, autochip, verilogeval, rtlfixer, rtllm, gpt4aigchip, rtlcoder}. Also for analog and mixed-signal IC design, several LLM-based workflows have been proposed: In \cite{analogcoder}, an LLM is provided with a custom SPICE interface to obtain simulation results, while other studies present workflows combining LLMs with custom circuit specific knowledge libraries \cite{ladac, ampagent, analogcoder, wiseeda} or with conventional optimization algorithms \cite{ladac, ampagent, ledro, wiseeda}. Furthermore, benchmarks have been established \cite{amsnet2, amsbench, masalachai} to evaluate LLM capabilities in analysis and design of analog and mixed-signal circuits.\\ 
While prior work has focused primarily on the development of ICs, this paper addresses the design of PCB modules based on switched-mode power supply (SMPS) circuits.
PCB module design poses unique challenges for applying LLMs. First, instead of being able to select component size parameters freely, PCB design requires the use of given catalogue-devices in specific packages, such as microcontrollers, power regulators, and passive components. A resulting concern is managing integration complexity, which involves numerous component pins, multiple voltage domains, diverse interfaces, and heterogeneous parts from many vendors. Secondly, in contrast to IC design, PCB designs are highly dependent on exact specifications and constraints from component datasheets and specific application notes. In addition, public data sources often lack critical information or access rights to certain documents can be restricted. As a result, it must be assumed that specific component information is often not part of an LLM's training dataset.\\
Furthermore, employing LLMs in circuit design faces other challenges. First, LLMs have a limited ability to interpret results from simulation tools like SPICE \cite{ladac}, which are essential in electronic development processes. As we demonstrate in Section \ref{subsec:LLM time series}, a multimodal LLM (GPT-4o) cannot reliably extract information from simulated time-series signals.
Second, many tasks require a complex, multi-step design process that involves information retrieval from datasheets, netlist analysis, netlist adaptation, solving equations, and integrating simulation feedback. Figure \ref{fig:thinking_steps} summarizes a reasoning chain required for an exemplary SMPS design task.

In this paper, we introduce different LLM-based workflows to address the described challenges and define a benchmark to investigate their effectiveness and limitations for switched-mode power supply (SMPS) circuit design.\\ 
To construct the LLM-based workflow, we develop a broad selection of domain-specific tools that serve as an interface between the LLM and the SPICE simulator. These functions reliably extract different features from the simulated signals and provide them as text-based feedback to the LLM, e.g. the peak-to-peak value of the output voltage ripple. The LLM conducts multiple iterations with the simulator and observes the effects of circuit modifications. Additionally, to enhance the LLM’s circuit-specific domain knowledge, it is enabled to interact with the corresponding datasheet via retrieval-augmented generation (RAG) \cite{rag}. Moreover, a state-of-the-art reasoning LLM is compared with a predecessor LLM that lacks reasoning capabilities.\\
To evaluate the LLM-based workflows, we conduct two experiments. The first experiment focuses on finding idealized component values, which is often an initial step in a design process. In the second experiment, the goal is to select an optimal set of catalogue components for circuit optimization.\\
For the first experiment, we create a comprehensive benchmark of 269 design tasks related to SMPS circuits, which are represented as SPICE netlists. Each task requires adapting a given circuit, see Figure \ref{fig:circuits}, to meet specific requirements. The benchmark covers circuits of varying complexity and different design task categories, such as topology adaption and parameter tuning. To test the LLMs’ capability to use simulation feedback, the benchmark contains tasks that explicitly require this feedback to compensate for imprecise datasheet equations or specifications. Furthermore, we investigate the abilities of the LLMs to solve complex multi-step SMPS design tasks, by varying the length and complexity of the required reasoning chain, see example reasoning chain in Figure \ref{fig:thinking_steps}.\\
In the second experiment, we test the ability of an LLM-based workflow to select components from a given set of catalogue components, to optimize the efficiency of an SMPS circuit in a specific operation mode.\\
In a comprehensive evaluation of the experiment results, the isolated contributions of the individual elements of the LLM-based workflow are presented and the performance over different design task categories is analyzed.
In addition, we discuss the computational cost and limitations of the proposed workflows to identify possible improvements, such as adapting the text-based representations of the SMPS circuits.\\
 
To summarize, this paper makes the following key contributions:
\begin{itemize}
    \item We present different LLM-based workflows for circuit design, which integrate a custom toolkit for interpreting SPICE simulation feedback, RAG for retrieving information from datasheets, and reasoning.
    \item In two comprehensive experiments, including a new benchmark with 269 design tasks, we evaluate capabilities and limitations of these workflows to assist in SMPS circuit design for PCB applications.
\end{itemize}

\section{Background}
\subsection{Modern Design of Switched-Mode Power Supplies}
Switched-mode power supplies (SMPS), such as the common Buck, Boost, and Buck-Boost topologies \cite{smps_report}, convert voltage levels with high efficiency. In a typical buck converter, a pulse-width modulation (PWM) controller rapidly switches a transistor to create a pulse train. This is subsequently filtered by an inductor and capacitor, see Figure \ref{fig:general_buck_converter}, to produce a DC output voltage whose level is determined by the transistor’s duty cycle.\\
Modern buck converters consist of dedicated ICs with advanced features. They enable different operation modes, such as burst and pulse-skipping mode, to achieve very high efficiencies and support spread spectrum operation to lower peak electromagnetic interference (EMI). Other advanced functionalities include adjustable startup times, adjustable switching frequencies, programmable thresholds for overvoltage (OVP) and overcurrent protection(OCP), and thermal shutdown. Buck converter topologies can also be extended to a multi-phase converters, where the ripple components partially cancel out each other due to phase shift, which reduces the output ripple.\\ 
Typical challenges in SMPS circuit design include component selection and sizing, topology design, efficiency and stability
considerations, while other important aspects like physical layout and thermal management are outside the scope of this work.

\subsection{LLMs and Time-Series Input}
\label{subsec:LLM time series}
Table \ref{tab:time series} provides the results of a case study, in which we examine the ability of a multimodal LLM (GPT-4o) to interpret time‑series data. Specifically, \mbox{GPT-4o} is tasked with identifying the ripple peak-to-peak value from either a raw numeric vector or its corresponding image representation. An answer is considered correct when the LLM's reading falls within a 10\% tolerance range of the ground-truth value.

\begin{table}[h]
\centering
\caption{Case study: Investigating the ability of the multimodal LLM GPT-4o to handle time series data, examined through the task of reading the ripple in time series  provided as numeric vectors or images}
\begin{tabular}{  m{1cm} | m{2cm}| m{2cm} | m{2cm}| m{0.5cm} | m{2cm}| m{0.5cm}} 
  \hline
  Test Case & Correct\newline Answer & Vector\newline Length & GPT-4o\newline (Vector) & & GPT-4o\newline (Image) & \\
  \hline
  1 & 426 µV & 1250 & 350 µV & {\Large \color{red}  \ding{55}}& 0.8 mV & {\Large \color{red}  \ding{55}} \\
  2 & 673 µV & 2040 & 605 µV & {\Large \color{green} \checkmark}& 1.1 mV & {\Large \color{red}  \ding{55}}\\
  3 & 17 mV & 1600 & 113 mV & {\Large \color{red}  \ding{55}}& 36 mV & {\Large \color{red}  \ding{55}}\\
  4 & 14.5 mV & 650 & 83.4 mV & {\Large \color{red}  \ding{55}}& 40 mV & {\Large \color{red}  \ding{55}}\\
  5 & 24.1 mV & 950 & 24.3 mV & {\Large \color{green} \checkmark}& 24 mV & {\Large \color{green} \checkmark}\\
  \hline
\end{tabular}
\label{tab:time series}
\end{table}

\begin{figure}[h]
    \centering
    \subfloat[Test case 1: Output voltage of buck converter with ripple]{
        \includegraphics[width=0.47\textwidth]{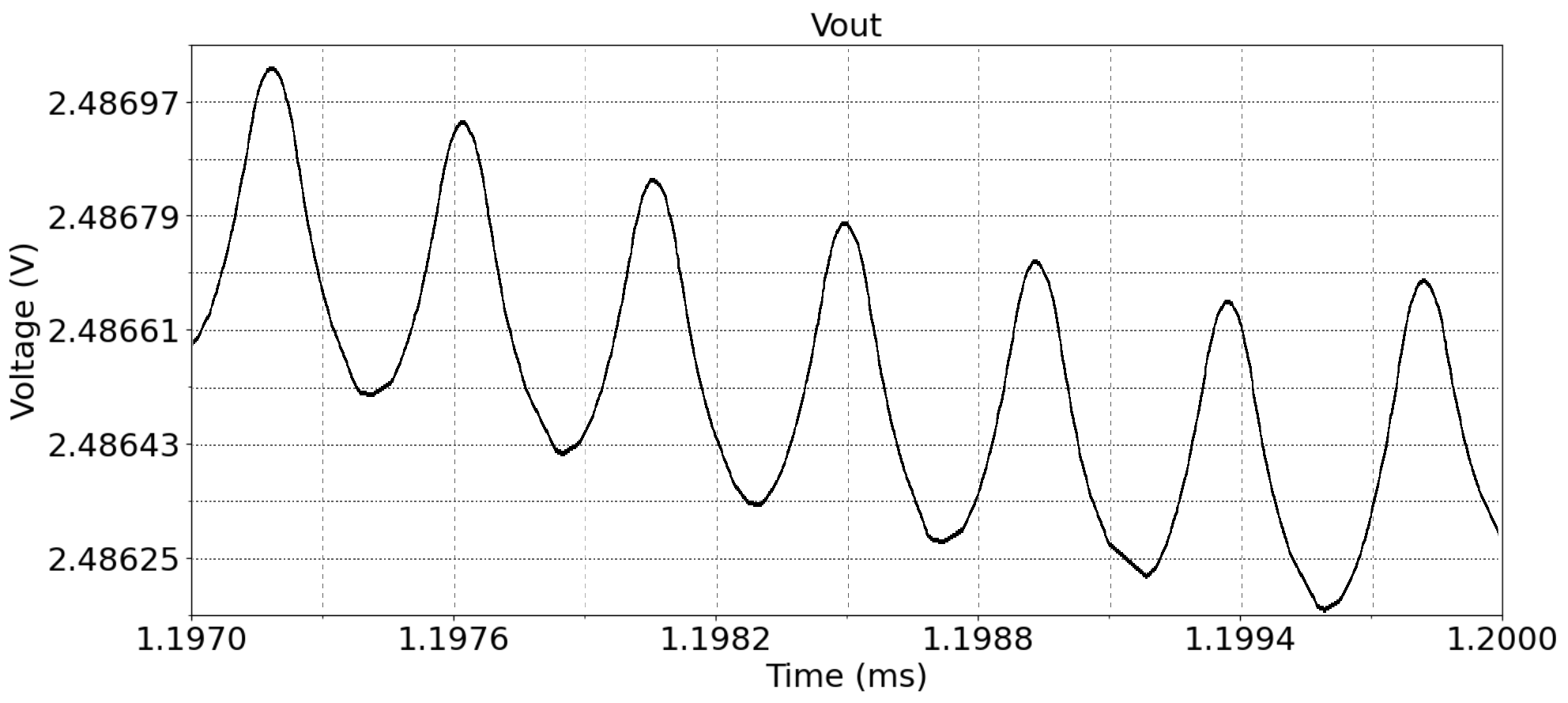}%
        \label{fig:ripple1}
    }
    \hfill
    %
    \subfloat[Test case 5: Output voltage of buck converter with ripple]{
        \includegraphics[width=0.47\textwidth]{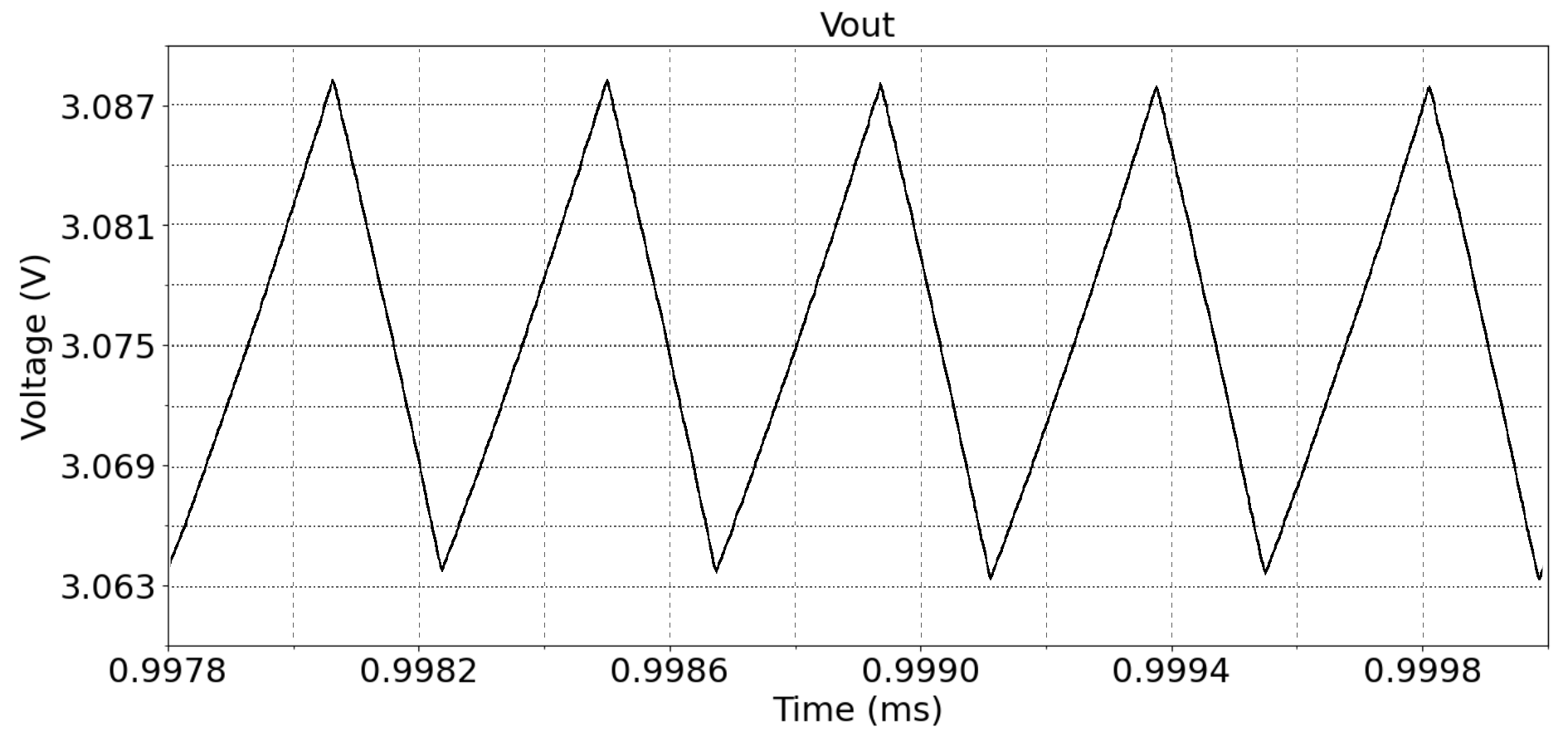}%
        \label{fig:ripple5}
    }
    \caption{Example SPICE simulation signals from the case study}
    \label{fig:ripple_signals}
\end{figure}

The results in table \ref{tab:time series} show, that it is not a reliable option to provide the time series as numeric vectors or images directly to the LLM for interpretation and further processing. 
While GPT-4o, for example, fails to extract the correct ripple from the signal in Figure \ref{fig:ripple1}, it succeeds with the test case depicted in Figure \ref{fig:ripple5}. The failure in Example \ref{fig:ripple1} may be due to the underlying trend in the ripple signal. While future multimodal LLMs might provide more precise answers, we must currently rely on rule-based extraction functions presented in the next section.

\subsection{Reasoning}
In this work, we evaluate two LLM models: OpenAI GPT-4o (released in 2024) and the newer LLM OpenAI o3 (released 2025). The key difference between the two models lies in their so-called reasoning capabilities.
Non-reasoning LLMs, like GPT-4o, generate answers directly based on pattern recognition and surface-level correlations in text. Reasoning LLMs additionally have the ability to simulate or represent internal deliberations before giving their final answer.\\
Reasoning techniques include chain-of-thought (CoT) prompting \cite{cot}, which instructs the LLM with examples how to generate intermediate reasoning steps before answering. Tree-of-thoughts (ToT) \cite{tree} expands on CoT by exploring multiple reasoning branches. Another reasoning technique, self-consistency decoding \cite{self}, runs multiple reasoning chains and selects the most consistent answer. Reasoning may also include an augmentation of LLMs with external tools, prompt rewriting, and further self-verification mechanisms.

\section{Methodology}
\subsection{LLM-assisted Simulation Workflows}
\begin{figure}[h]
  \centering
  \includegraphics[width=\linewidth]{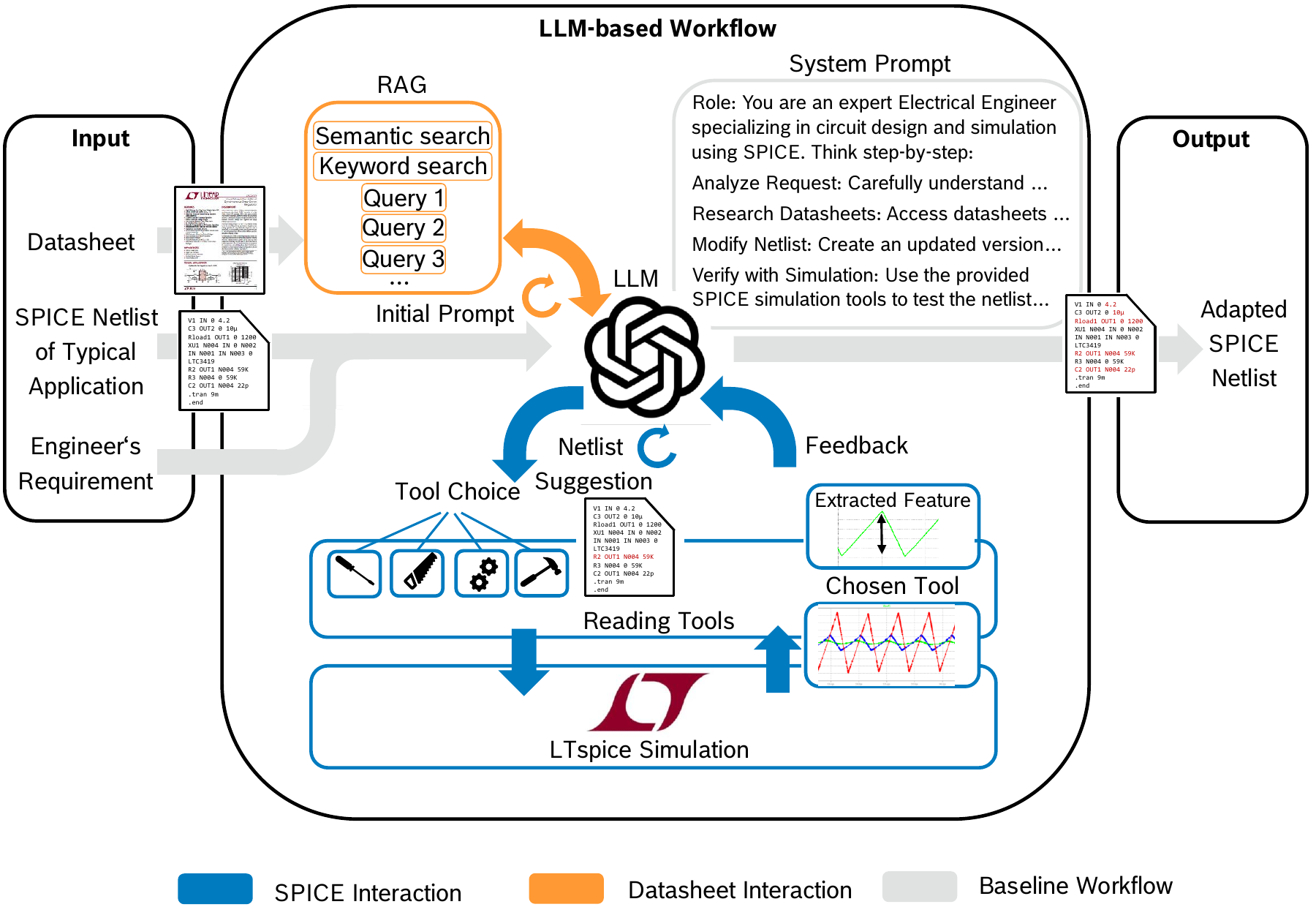}
  \caption{Overview of the LLM-based workflows: The baseline workflow uses an instruction-tuned LLM (GPT-4o or o3) that receives a reference circuit and requirements as input to directly generate an adapted circuit netlist. The first extension enhances this workflow with information retrieval from datasheets using RAG. The second extension equips the LLM-based workflow with a comprehensive set of feature extraction tools to receive simulation feedback}
  \label{fig:architecture}
\end{figure}
Figure \ref{fig:architecture} introduces the structure of the LLM-based workflows discussed in this paper. The model receives the netlist of a typical application circuit from the SMPS controller's datasheet as input.
In most cases these initial circuits do not meet the desired specifications. Therefore, the main goal is to modify the netlist to align with specific input requirements given by design engineers.\\
In the setup shown in Figure \ref{fig:architecture}, the standalone LLM serves as baseline, depicted in gray, which directly outputs a modified netlist. 
To optimize the LLMs performance, we employ prompt engineering, providing a tailored system prompt. For instance, the LLM agent is instructed to always include the adapted netlist in its responses.\\ 
The first extension, depicted in orange, enables the LLM to use RAG to retrieve specific information from the corresponding SMPS controller datasheet. The RAG system is accessed via an API call and can perform both keyword and semantic searches in vector stores and running multiple queries in parallel. The queries are generated by the LLM based on its input and system prompt. The LLM autonomously defines the number of needed RAG calls.\\ 
The second workflow extension, highlighted in blue in Figure \ref{fig:architecture}, allows the LLM agent to receive simulation feedback on the circuit's behavior. This feedback enables the LLM to estimate the impact of netlist modifications and verify suggested adaptations, thereby enhancing response reliability.
This interaction is enabled by a set of reading tools (Python functions) that serve as an interface between the LLM and the SPICE simulation. As noted in Section \ref{subsec:LLM time series}, multimodal LLMs like GPT-4o cannot reliably extract features, such as output ripple, directly from simulation signals. To overcome this limitation, a comprehensive set of tools is developed to extract these features and return them to the LLM in a text-based format.
The LLM is provided with a list of JSON files, each documenting a tool's functionality, name, and required arguments. Based on this information, the LLM autonomously decides which tool to use. Similarly it determines the number of API tool calls, depending on the specific benchmark question and the history of previous tool feedback. Examples of these tools include:
\begin{itemize}
    \item \textit{get\_mean\_output\_voltage()}\\
    Calculates the mean over a time interval in the steady state of the output voltage.
    \item \textit{get\_ripple()}\\
    Extracts the peak-to-peak value of a periodic signal in the steady state, as discussed in Section \ref{subsec:LLM time series}.
    \item \textit{get\_switching\_frequency()}\\ 
    First, it performs a Fourier transform of the signal within a steady-state time interval, then extracts the fundamental frequency by identifying the highest peak in the spectrum.
    \item \textit{get\_startup\_time()}\\
    Calculates the mean value of the output voltage over a time window in steady state and identifies the first time the signal reaches 90\% of the steady state value.
\end{itemize}

\subsection{Benchmarking}
\label{subsection:benchmarking}
\begin{figure}[h]
    \centering

    \subfloat[Textbook buck converter circuit in LTSpice (easy)]{
        \includegraphics[width=0.42\linewidth]{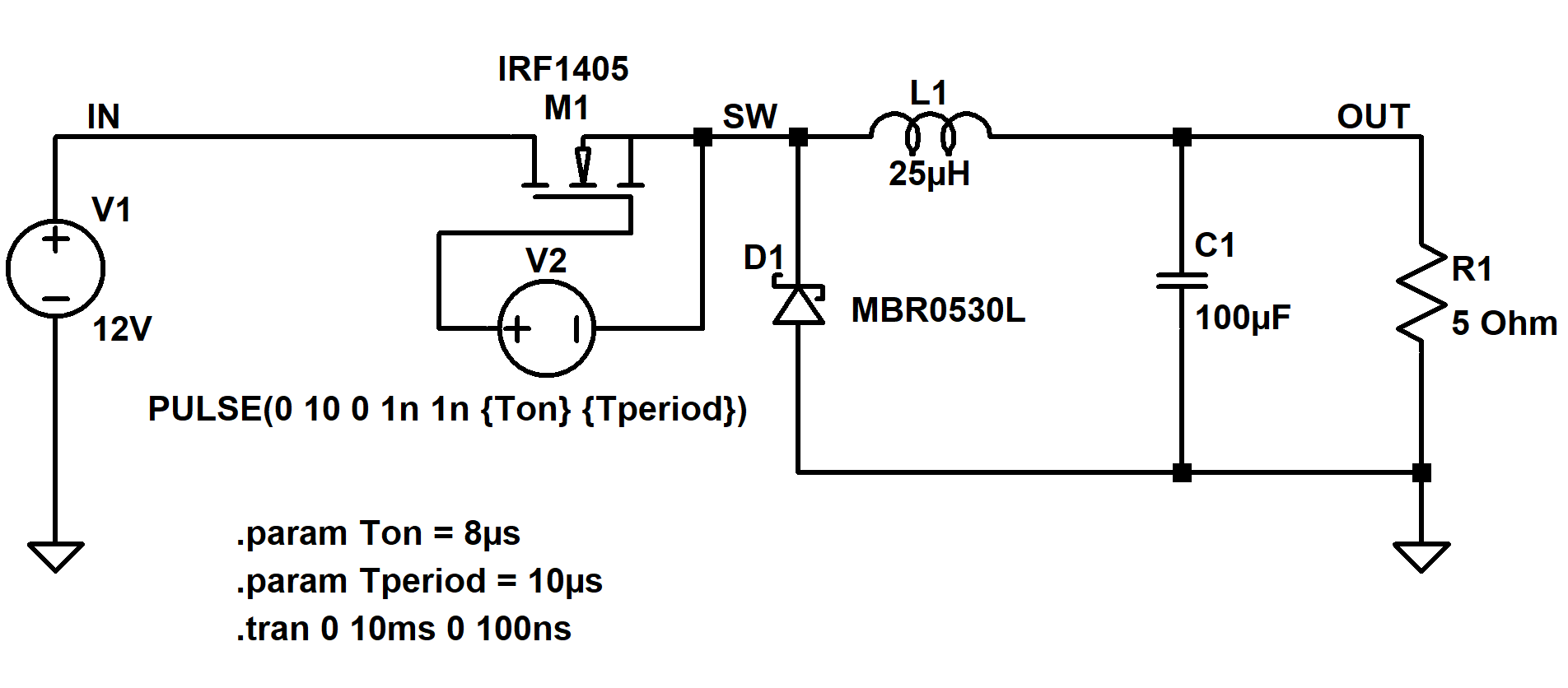}
        \label{fig:general_buck_converter}
    }
    \hfill
    \subfloat[Typical LTC3419 application circuit in LTSpice, after \cite{ltc3419} (medium)]{
        \includegraphics[width=0.42\linewidth]{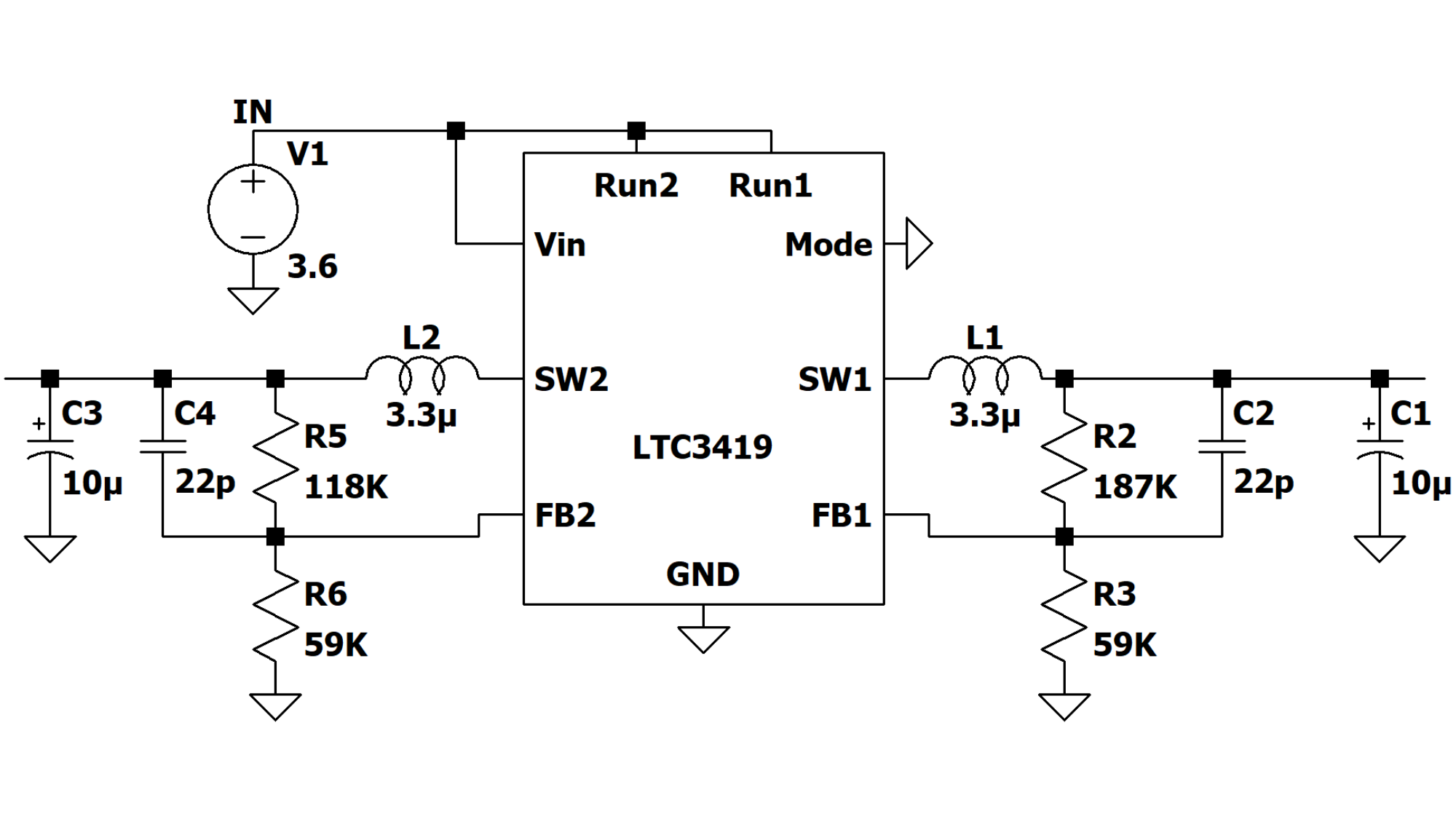}
        \label{fig:ltc3419}
    }

    \subfloat[Typical LTC7802 application circuit in LTSpice, after \cite{ltc7802} (hard)]{
        \includegraphics[width=0.85\linewidth]{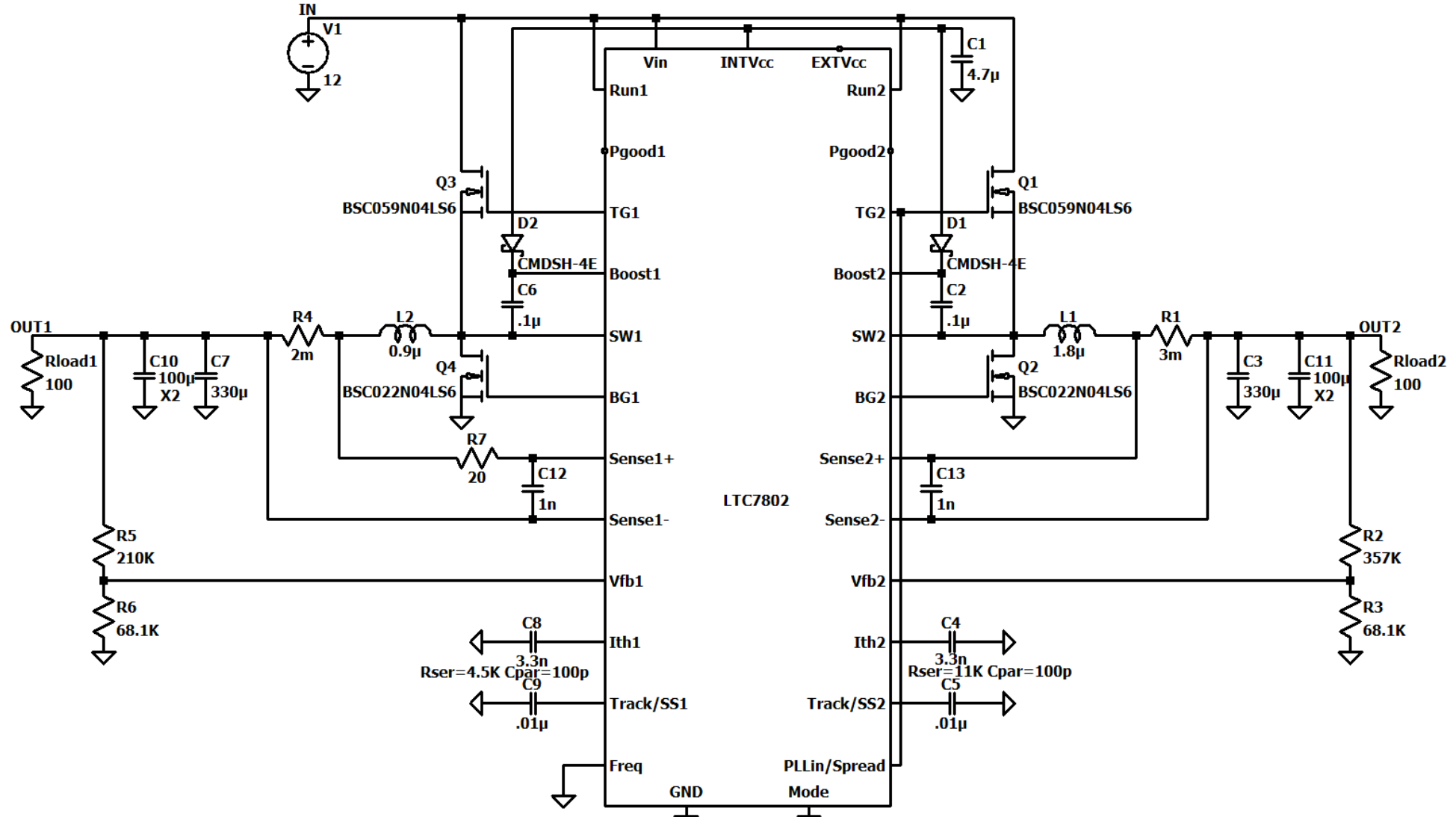}
        \label{fig:ltc7802}
    }

    \caption{The three SMPS circuit types that serve as the basis for the benchmark, shown in order of increasing difficulty}
    \label{fig:circuits}
\end{figure}

\begin{figure}[h]
  \centering
  \includegraphics[width=\linewidth]{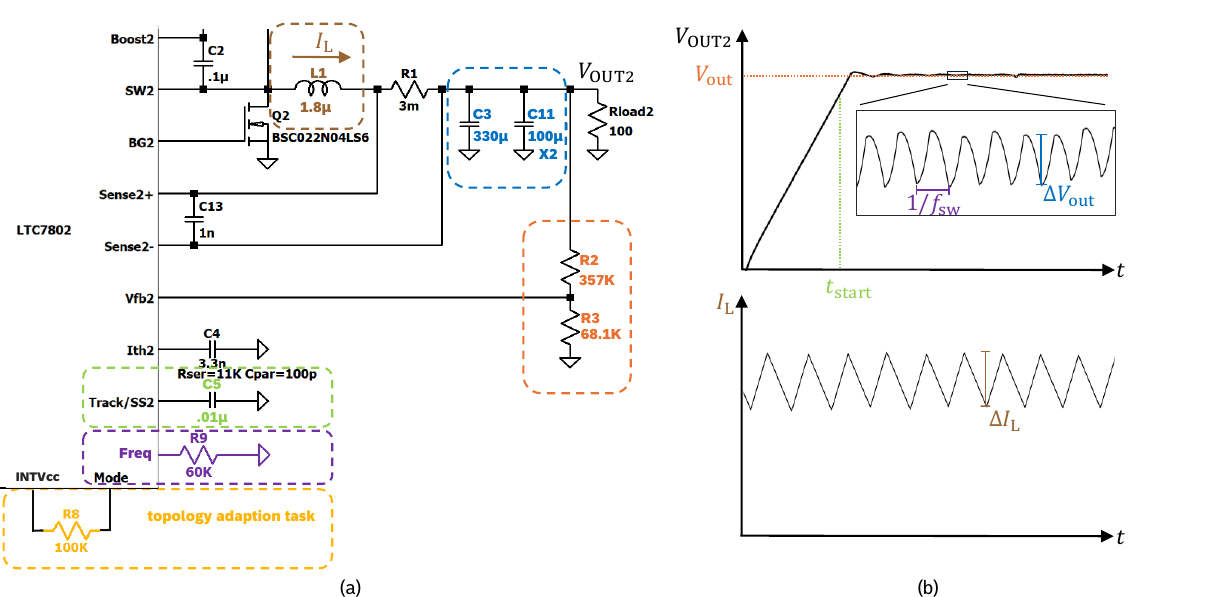}
  \caption{Illustration of a topology adaption task and the different parameter tuning tasks - $V_{\mathrm{out}}$, $\Delta I_{L}$, $\Delta V_{\mathrm{out}}$, $f_{\mathrm{sw}}$, $t_{\mathrm{start}}$. For the different task categories affected elements in the circuit excerpt (a) and in the simulation signal segments (b) are marked in the same color.}
  \label{fig:feature_explanations}
\end{figure}

We developed a benchmark to evaluate the ability of LLM-based workflows to assist in the circuit design process of SMPS. The benchmark tasks are based on SMPS example circuits, featuring three circuit types with increasing levels of complexity. The most simple circuit is a textbook buck converter, illustrated in Figure \ref{fig:general_buck_converter}. The medium level is a buck converter using the LTC3419, a dual step-down regulator \cite{ltc3419}, as shown in Figure \ref{fig:ltc3419}. The most complex example is the typical application circuit of the LTC7802,
a 2-phase synchronous step-down controller with advanced features like spread spectrum operation or a programmable switching frequency, depicted in Figure \ref{fig:ltc7802}.\\
Overall, the benchmark contains 269 test questions: 72 for the textbook buck converter (easy), 72 for the LTC3419 (medium), and 125 for the LTC7802 (hard). 
The benchmark can be divided into two design task categories: \textit{parameter tuning} and \textit{topology adaption}. The parameter tuning tasks can  further be subdivided according to the specific parameter being adjusted: supply voltage $V_{\mathrm{in}}$, output voltage $V_{\mathrm{out}}$, current ripple $\Delta I_{L}$, output voltage ripple $\Delta V_{\mathrm{out}}$, and for the LTC7802 only, the switching frequency $f_{\mathrm{sw}}$ and startup time $t_{\mathrm{start}}$. Figure \ref{fig:feature_explanations} illustrates required circuit modifications and the resulting effects in the output voltage and inductor current for each design task category.

Solving the $V_{\mathrm{out}}$-, $\Delta I_{L}$- and $\Delta V_{\mathrm{out}}$- tuning tasks requires the LLM to process a reasoning chain whose complexity and length increase from task to task. To explain this, we briefly describe the necessary steps to solve these tasks.\\
To tune the output voltage $V_{\mathrm{out}}$ in LTC3419 or LTC7802 circuits, the following equation 
\begin{equation}
  V_{\mathrm{out}} = V_{\mathrm{ref}}\left(1+\frac{R_{\mathrm{top}}}{R_{\mathrm{bot}}}\right)
  \label{eq:vout}
\end{equation}
and the corresponding reference voltage $V_{\mathrm{ref}}$ must be retrieved from the datasheet, rearranged and calculated to receive the values for $R_{\mathrm{top}}$ or $R_{\mathrm{bot}}$. These feedback resistors must be identified in the netlist and adapted accordingly. Additional factors, such as ensuring sufficient supply voltage, must be considered.\\
For current ripple $\Delta I_{L}$ adjustment tasks, the following equation
\begin{equation}
  \Delta I_{L} = \frac{1}{f_{\mathrm{sw}}L}V_{\mathrm{out}}\left(1-\frac{V_{\mathrm{out}}}{V_{\mathrm{in}}}\right)
  \label{eq:deltaI}
\end{equation}
must be retrieved, rearranged, and used to calculate the required value for the inductor $L$. To obtain the missing parameters for Equation \ref{eq:deltaI}, all steps from the previous example must be repeated to get the output voltage. Additionally, input voltage and switching frequency must be identified from the datasheet and the circuit configuration in the netlist.\\
For solving output voltage ripple $\Delta V_{\mathrm{out}}$ tasks, first Equation \eqref{eq:vout} and \eqref{eq:deltaI} must be solved again and the corresponding results used to determine the required $ESR$ with the equation  
\begin{equation}
  \Delta V_{\mathrm{out}} \approx \Delta I_{L} \left(ESR + \frac{1}{8f_{\mathrm{sw}}C_{\mathrm{out}}}\right).
  \label{eq:deltaV}
\end{equation}

Typically, for \textit{parameter tuning} tasks, the LLM assistant must perform several key steps: 
\begin{itemize}
    \item Solve equations
    \item Retrieve information from datasheet and circuit configuration
    \item Identify and adapt the relevant components in the netlist
\end{itemize}
The equations to determine $\Delta V_{\mathrm{out}}$ and $t_{\mathrm{start}}$ are only approximations. Consequently, these tasks specifically test the LLM's ability to integrate simulation feedback.\\
The second category, \textit{topology adaption}, includes tasks such as inserting key components into the converter topology (for example diode, inductor and MOSFETs), adjusting controller channel configurations, enabling or disabling spread spectrum operation, or selecting a specific operating mode. For example, to enable pulse skipping mode for the LTC7802 controller, the peripheral circuit must be adjusted by connecting the MODE pin to the INTVcc pin through a high-impedance resistor. The LLM can collect this information by accessing the datasheet via RAG. It would then identify these two pins in the netlist and modify the circuit accordingly.\\

\begin{figure}[h]
  \centering
  \begin{minipage}[b]{0.42\linewidth}
    \centering
    \includegraphics[width=\linewidth]{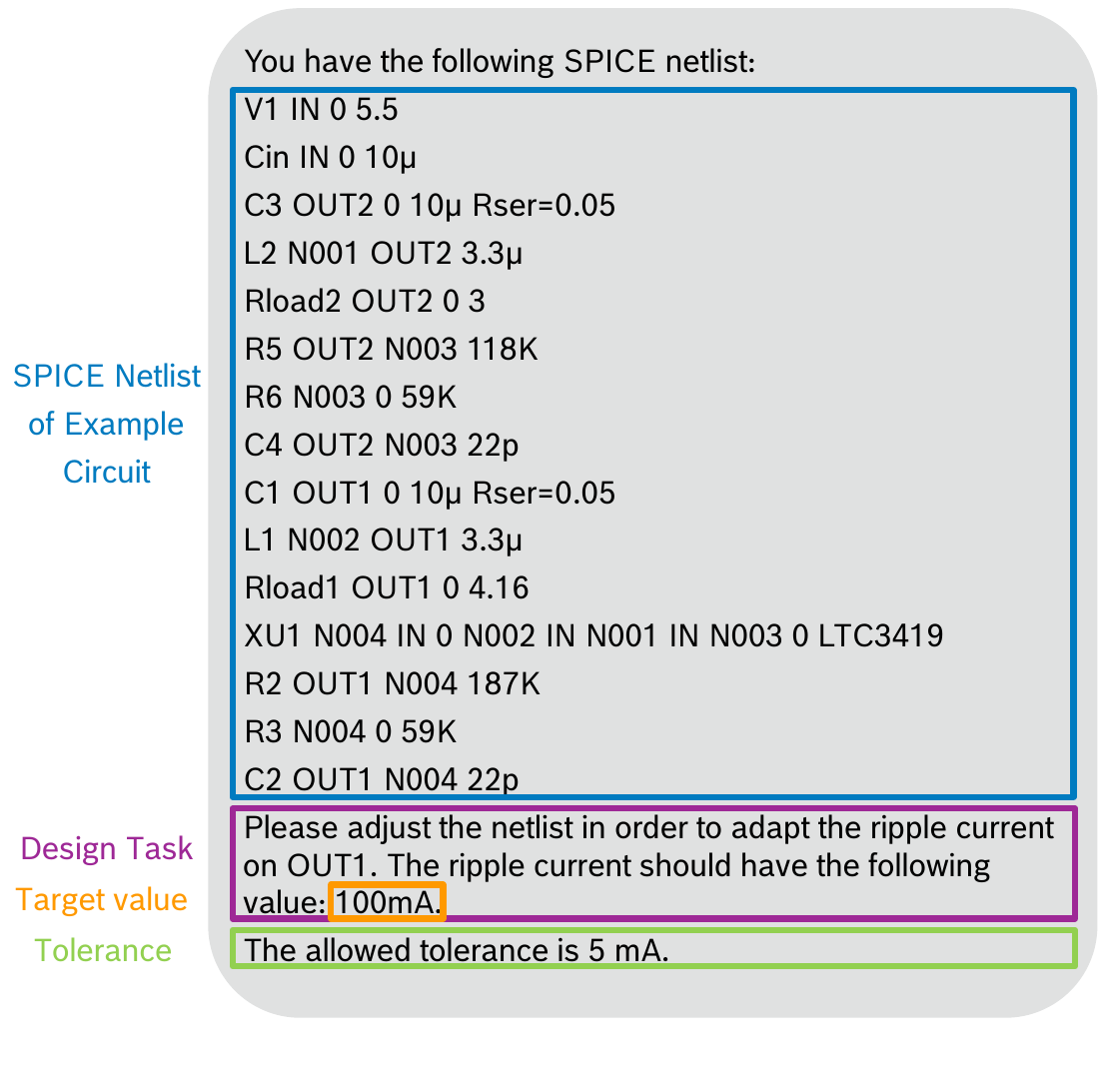}
    \caption{Exemplary benchmark input prompt}
    \label{fig:colored_prompt}
  \end{minipage}
  \hspace{0.02\linewidth}
  \begin{minipage}[b]{0.53\linewidth}
    \centering
    \includegraphics[width=\linewidth]{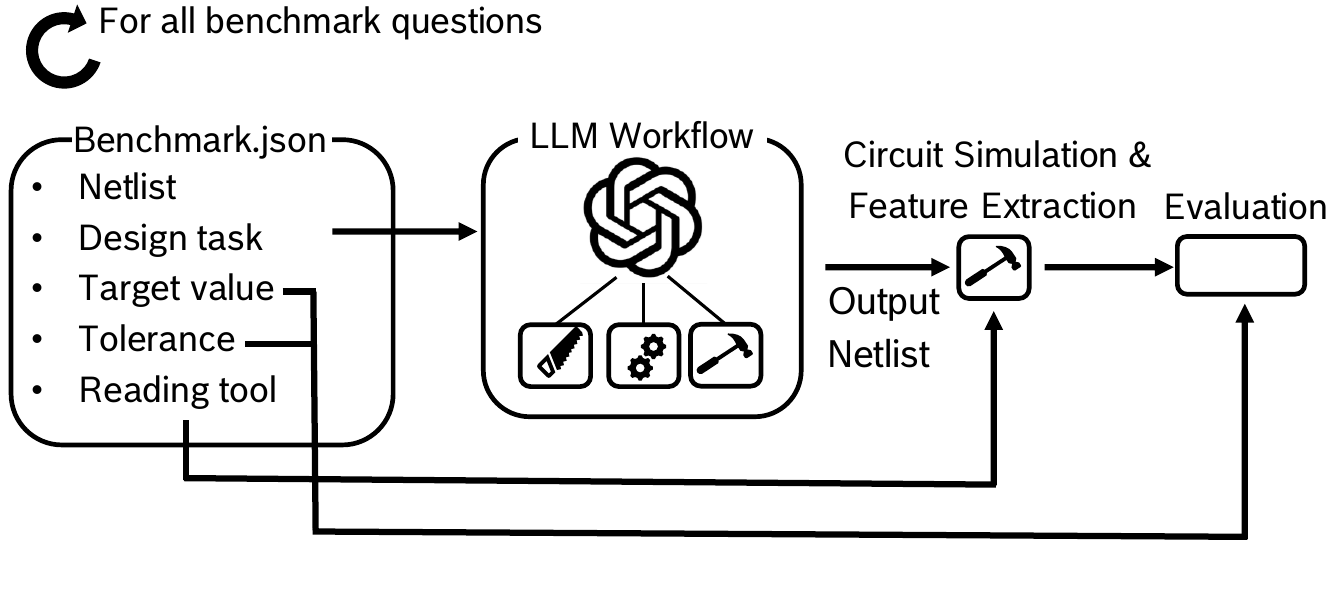}
    \caption{The automatic benchmarking involves simulating the circuit received by the LLM workflow, extracting the relevant design feature from the simulation using a reading tool, and evaluating it against the target value within the specified tolerance}
    \label{fig:automatic_benchmarking}
  \end{minipage}
\end{figure}

The 269 benchmark questions were created by combining different design tasks, varying component values and the three circuit types. Further data diversification is achieved by varying the appearance of the netlist by deliberately changing node and component names and applying minor topology changes. Moreover, integrating the design tasks with different target parameters, such as setting the ripple current to different levels, increases the number of benchmark questions. A complete exemplary benchmark prompt, using a medium-difficulty LTC3419 netlist as input for the LLM-based workflows is shown in Figure \ref{fig:colored_prompt}.\\ 
The benchmarking process is automated, as visualized in Figure \ref{fig:automatic_benchmarking}, where the same reading tools can be applied in the LLM-based workflow and the evaluation.\\
In our first experiment, the LLM is tasked to find ideal component values, which can be used subsequently to find suitable commercially available parts. The second experiment, described in Section \ref{subsection:real_components}, evaluates the LLM workflow's performance in selecting real catalogue-components.\\

\section{Experiment Results}
\subsection{Experimental Setup}
The evaluated LLMs, OpenAI GPT-4o and o3, are accessed via the Azure OpenAI REST API \cite{azure_api} (version 2025-04-01-preview) using the responses and assistant interface. The datasheet of the corresponding controller serves as database for the RAG system. The RAG process is triggered via API tool calls by the LLM and executes keyword as well as semantic searches in vector stores. For the vector embedding the text-embedding-3-large model \cite{embedding} with 256 dimensions is applied. The maximum number of chunks added as context to the LLM is 20, with each chunk having the size of 800 tokens.
 
Circuit simulations are performed using the simulator LTSpice \cite{LTSpice}. All reported API call results were generated between March and September 2025.

\subsection{Evaluation Metrics}
As evaluation metric the solve rate is used, which represents the percentage of design tasks from the benchmark that are correctly completed. A question is considered to be correctly answered if the target value falls within a specified tolerance range, such as a ripple of 18~mV~$\pm$~0.9~mV. Generally, the tolerance is set at 5\%, except for topology adaption questions, such as "Adjust the netlist to select the pulse skipping mode.", which can only be answered right or wrong.\\ 
As supplementary continuous metric we use the absolute percentage error ${APE= \left| \frac{A_i - F_i}{A_i} \right|\cdot 100\%}$, where $A_i$ represents the actual value and $F_i$ the value provided by the LLM agent. Due to the presence of significant outliers in the $APE$ we use the median of the $APE$ to avoid distortion. 
For the median $APE$, topology adaption questions are excluded since they can only be marked as correct or incorrect.\\ 
To receive a better statistical estimation of the performance and to determine what constitutes a statistically significant change in the performance metrics, we run the entire benchmarking process $n$ times, with $n=40$. This allows us to calculate the sample mean of the evaluation metrics and the corresponding confidence interval \cite{statistic} with a significance level of $\alpha=0.05$. Due to limitations concerning the simulation runtime, this statistical evaluation is only performed for benchmark questions targeted on the simple circuit type of the textbook buck converter. 

\subsection{Experiment 1: Parameter Tuning and Topology Adaption with Idealized Components}
The benchmarking is conducted for the following workflow configurations: GPT-4o and o3 standalone, combined with datasheet interaction ( + RAG), with simulation feedback ( + SPICE) and both. The benchmarking results are presented in Table \ref{tab:main}. The boxplot in Figure \ref{fig:ape_boxplot} illustrates the distribution of the $APE$ across the entire benchmark, using the reasoning LLM o3 for all four tested approaches.\\
\begin{table}[h]
\caption{Experiment 1 - Benchmarking results: The performance is shown for design tasks related to the three different circuit types and the total benchmark, using the solve rate and median($APE$) as performance metrics. The methods assessed include GPT-4o and o3 agents, combinations with datasheet information retrieval via RAG and feedback from SPICE simulations
}

\footnotesize
\begin{adjustbox}{max width=\textwidth}
\begin{tabular}{l|l|l|l|l|l|l||l|l}

    \hline
      & \multicolumn{2}{>{\centering\arraybackslash}p{3cm}|}{Textbook Buck Converter (easy)} & \multicolumn{2}{c|}{LTC3419 (medium)} & \multicolumn{2}{c||}{LTC7802 (hard)} & \multicolumn{2}{c}{Total}  \\
    \hline
     Method & Solve Rate & med($APE$) & Solve Rate & med($APE$) & Solve Rate & med($APE$) & Solve Rate & med($APE$) \\
    \hline
     GPT-4o & 29.0 $\pm$ 1.0*& 55.3 $\pm$ 3.1* & 12.5 & 56.3 & 7.1 & 75.2 & 14.4 & 64.8 \\
     o3** & 44.9 $\pm$ 1.2*& 13.8 $\pm$ 1.0* & 18.1 & 32.8 & 18.4 & 31.7 & 25.4 & 27.2 \\
     \hline
     GPT-4o + RAG & 25.4 $\pm$ 1.1* & 63.2 $\pm$ 2.7* & 18.1 & 55.0 & 9.6 & 59.6 & 16.1 & 59.3 \\
     o3 + RAG** & 58.0 $\pm$ 1.0*& 3.1 $\pm$ 0.2* & 44.4 & 12.0 & 36.0 & 8.1 & 44.1 & 7.8 \\
     \hline
     GPT-4o + SPICE & 64.4 $\pm$ 1.5* & 3.7 $\pm$ 0.2* & 48.6 & 4.7 & 35.4 & 7.2 & 46.4 & 5.6 \\
     o3 + SPICE** & \textbf{96.5 $\pm$ 0.6}*& \textbf{1.6 $\pm$ 0.1} & \underline{95.8} & \textbf{0.8} & \underline{72.0} & \textbf{1.4} & \underline{84.9} & \textbf{1.3} \\
     \hline
     GPT-4o + RAG + SPICE &  64.9 $\pm$ 1.8* &  3.6 $\pm$ 0.3* &  50.0 &  4.6 & 42.4 & 4.6 & 48.7 & 4.4 \\
     o3 + RAG + SPICE** &  \textbf{97.2 $\pm$ 1.3*} &  \textbf{1.4 $\pm$ 0.2*} &  \textbf{97.2} &  \underline{0.9} &  \textbf{83.2} &  \underline{1.7} & \textbf{90.7} & \underline{1.4} \\
     
  \hline

\end{tabular}
\end{adjustbox}
\label{tab:main}
\vspace{2mm}
\parbox{\textwidth}{
*The whole benchmarking is performed n=40 times to receive confidence intervals, **Agents using reasoning}
\end{table}
Table \ref{tab:main} shows that the standalone LLMs achieve total solve rates of 14.4\% (LLM GPT-4o) and 25.4\% (reasoning LLM o3), respectively. In the workflow combined with RAG, the total solve rate increases to 16.1\% (LLM GPT-4o) and 44.1\% (reasoning LLM o3), respectively. When simulation feedback is integrated into the workflow, GPT-4o solves 46.4\% and o3 84.9\% of all benchmarking tasks. When using both SPICE and RAG, the performance of GPT-4o increases to 48.7\%. The best solve rate of 91\% is achieved by the o3 reasoning model combined with SPICE feedback and RAG.
From these results, we can conclude that the two main factors improving performance are SPICE simulation feedback and reasoning capabilities of the LLM.

\subsubsection{SPICE simulation feedback:}
\begin{figure}[h]
  \centering
  \begin{minipage}[t]{0.45\linewidth}
    \centering
    \includegraphics[width=\linewidth]{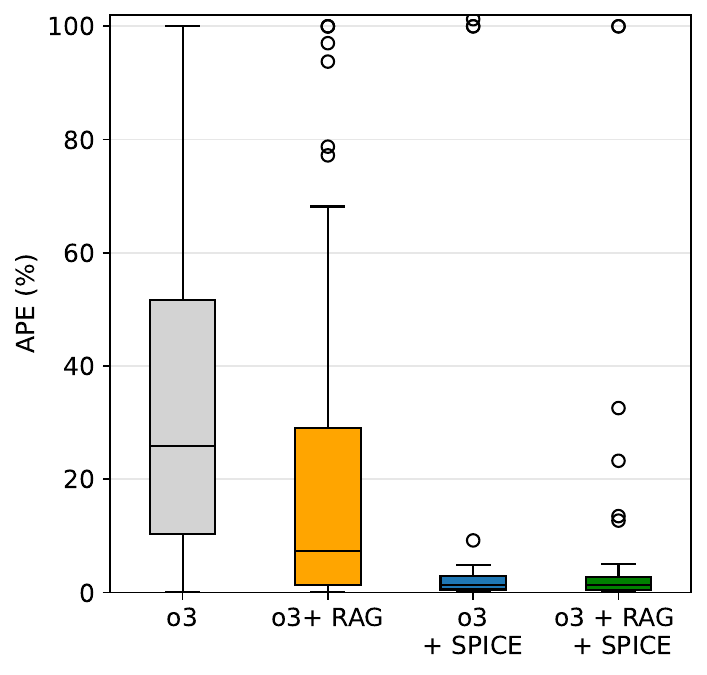}
    \caption{The distribution of the $APE$ on the total benchmark for the four tested approaches: o3, o3 + RAG, o3 + SPICE and o3 + RAG + SPICE. The central box depicts the range from the first quartile (Q1) to the third quartile (Q3), with a line marking the median} 
    \label{fig:ape_boxplot}
  \end{minipage}
  \hspace{0.02\linewidth}
  \begin{minipage}[t]{0.45\linewidth}
    \centering
    \includegraphics[width=\linewidth]{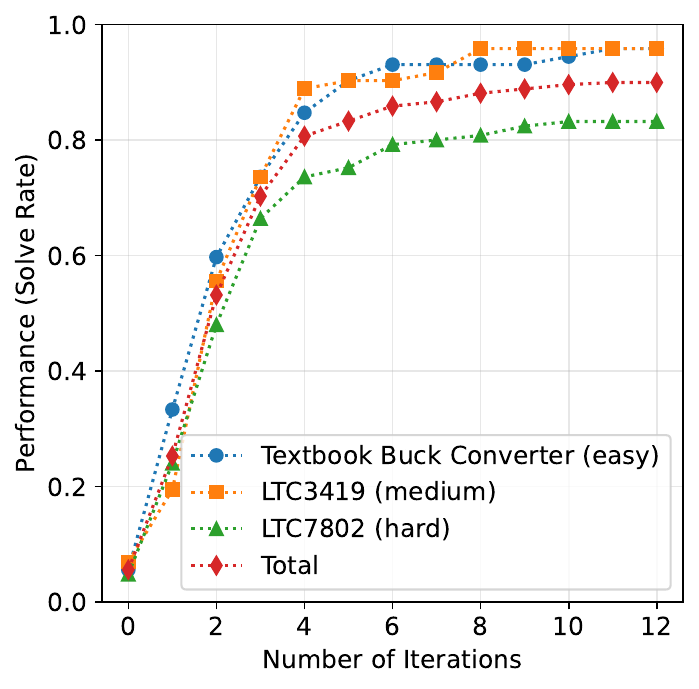}
    \caption{Performance of the LLM workflow o3 + RAG + SPICE over the number of tool iterations for the circuit types textbook buck converter (easy), LTC3419 (medium), LTC7802 (hard) and the total benchmark}
    \label{fig:tool_iterations}
  \end{minipage}
\end{figure}

Table \ref{tab:main} shows that the total solve rate of the standalone LLMs significantly increases through integrating SPICE feedback (LLM vs. LLM + SPICE) by 32\% (LLM GPT-4o) and 59.5\% (reasoning LLM o3). The simulation feedback improves the solve rate for all circuit complexities. Similarly, in Figure \ref{fig:ape_boxplot} it can be seen in the boxplot that SPICE feedback reduces the mean and variance of the $APE$.\\
This shows that the LLM effectively uses the offered simulation tools.\\
The simulation feedback increases the solve rate in all cases because it enables the LLM to verify its choices and correct previous mistakes. A common error are wrong assumptions during formula solving. For instance, using an incorrect $C_{\mathrm{out}}$ for voltage ripple calculation (Equation \eqref{eq:deltaV}) by failing to recognize all output capacitors from the netlist, or retrieving the wrong reference voltage from the datasheet for output voltage adjustments (Equation \eqref{eq:vout}). Such errors can be observed in the simulation response and iteratively fixed by the LLM .\\
Another possible source of mistakes is that the LLM agent fails to follow its instructions. SPICE feedback ensures it follows instructions more closely, minimizing errors such as returning incomplete netlists.\\
In some cases, the LLM lacks critical information. With simulation feedback, it can still achieve good results through initial estimates and iterative refinement, demonstrated by the performance for the o3 + SPICE workflow 
without datasheet access.\\
In Figure \ref{fig:tool_iterations}, the interaction of the best performing workflow (o3 + RAG + SPICE) with the simulation tools is analyzed. The performance measured by the solve rate is shown over the number of simulation feedback iterations. We observe a rapid increase in the solve rate within the first four iterations until a plateau is reached. This demonstrates that a small number of simulation iterations ($\leq$4) are sufficient to achieve a significant performance improvement.

\subsubsection{Reasoning:}
According to Table \ref{tab:main}, reasoning improves the performance of the standalone LLM by 9\%, with RAG (LLM + RAG) it improves by 28\%, with simulation feedback (LLM + SPICE) by 38.5\% and in combination (LLM + SPICE + RAG) by 42\%.\\ 
These observations reveal that reasoning is beside simulation feedback a main factor that enhances the performance of all workflows. Further they show that both SPICE feedback and RAG are used much more effectively by the reasoning model.\\ 
The LLM GPT-4o tends to complete substeps of multi-step benchmark tasks insufficiently, because it lacks reasoning capabilities.
For example, it solves formulas incorrectly or does not adapt the circuit netlist. It is also prone to making false assumptions, such as guessing a $V_{\mathrm{out}}$ of 5V without calculating it within the task to determine the current ripple with Equation~\eqref{eq:deltaI}.\\
In contrast, reasoning LLMs explore a broader range of possible solution paths. For example, in a task to find a maximum possible ESR to ensure a maximal voltage ripple, the non-reasoning model ignores two of three cascaded output capacitors, while the o3 reasoning model correctly includes them in the calculation of Equation~(\ref{eq:deltaV}) and the subsequent component adjustments.

\subsubsection{Retrieval-Augmented Generation:}
When combining RAG and the non-reasoning model GPT-4o, Table \ref{tab:main} shows only a slight improvement in the total solve rate by 1.7\%. Results are different for the reasoning LLM o3, where using RAG to retrieve information from the datasheet significantly increases the total solve rate from 25.4\% to 44.1\% (o3 vs. o3 + RAG) and from 84.9\% to 90.7\% (o3 + SPICE vs. o3 + RAG + SPICE).\\
This means that LLMs, which lack reasoning capabilities, are less able to integrate complex retrieved information into their problem-solving workflow. We confirm this by an additional small experiment. The LLM GPT-4o is evaluated on our benchmark using an "idealized RAG" for the textbook buck converter tasks by manually including all relevant information in the input prompt. However, these results do not surpass those achieved with the standard RAG.

\subsubsection{Increasing circuit complexity:}
The performance of the tested workflows in Table \ref{tab:main} declines from the textbook buck converter (easy) to the LTC3419 (medium) and further to the LTC7802 (hard). This trend can be attributed to several possible reasons. First, the greater amount of publicly available data about the general textbook buck converter compared to the specific controllers LTC3419 and LTC7802. Second, the
LTC3419, and especially the LTC7802, have more complex netlists with more components and nodes, an increasing number of controller pins to configure. Therefore, these text-based circuit netlists are increasingly difficult to comprehend and leave more possibilities for mistakes compared to the significantly shorter netlist of the textbook buck converter.\\

\begin{figure}[h]
  \centering
  \includegraphics[width=\linewidth]{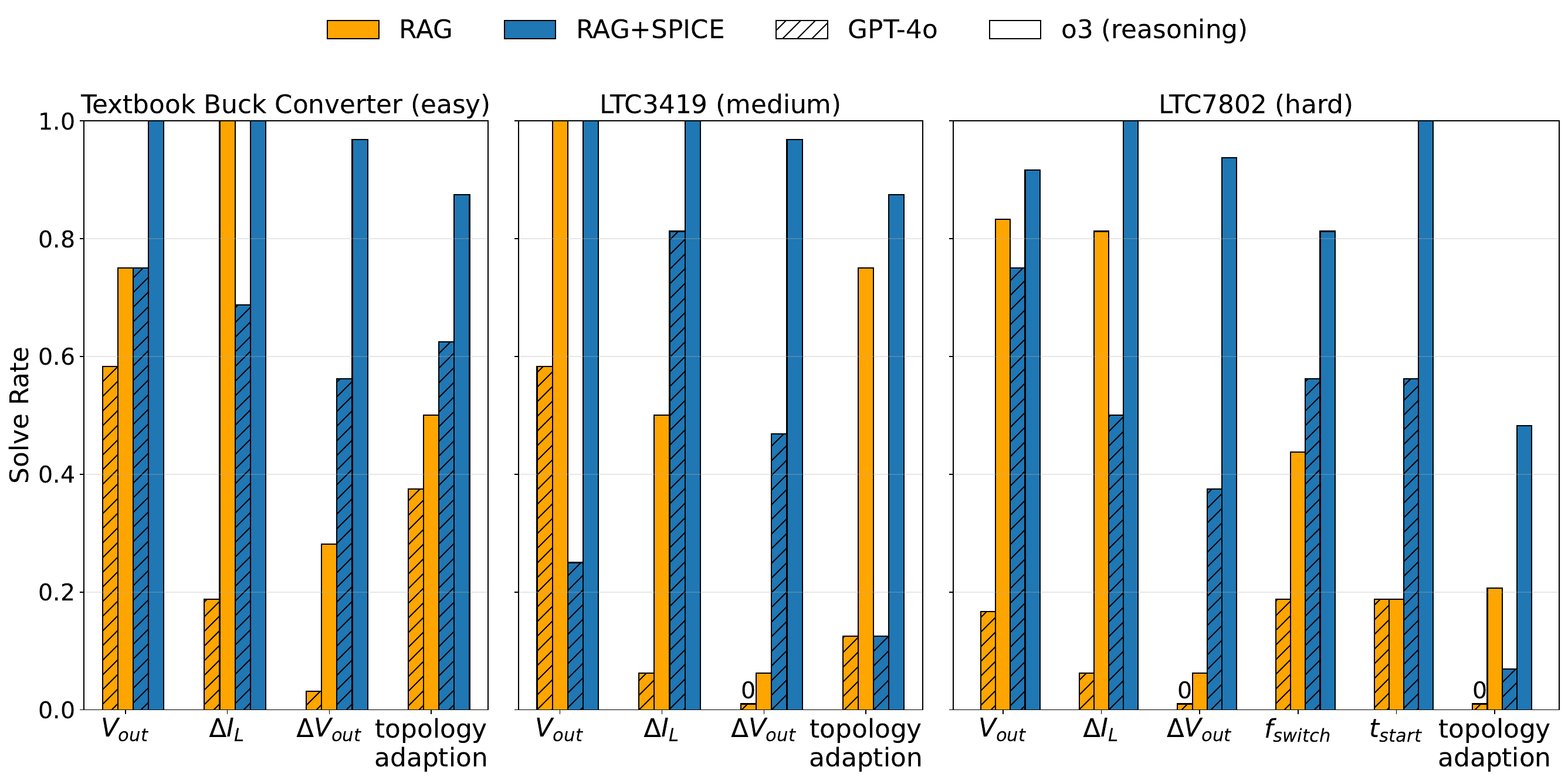}
  \caption{The solve rate of several LLM-based workflows split by question category. The parameter tuning design tasks are subdivided into categories according to the target parameters: output voltage $V_{\mathrm{out}}$, ripple current $\Delta I_{L}$, output voltage ripple $\Delta V_{\mathrm{out}}$, switching frequency $f_{\mathrm{switch}}$ and the startup time of the controller $t_{\mathrm{start}}$. Shown are LLM workflows using GPT-4o and the o3 reasoning model combined with RAG for information retrieval from the corresponding datasheet and SPICE simulation feedback}
  \label{fig:question_categories}
\end{figure}

\subsubsection{Analysis of design task categories:}
Section \ref{subsection:benchmarking} introduced the design task categories of topology adaption and parameter tuning. The latter can be further subdivided by the specific parameter being adjusted: output voltage $V_{\mathrm{out}}$, current ripple $\Delta I_{L}$, output voltage ripple $\Delta V_{\mathrm{out}}$, and, for LTC7802 only, the switching frequency $f_{\mathrm{sw}}$ and startup time $t_{\mathrm{start}}$. Figure \ref{fig:question_categories} presents the solve rates of different LLM workflows, broken down by these task categories and the three circuit types.\\
The best workflow - o3 (reasoning LLM) + RAG + SPICE - solves most parameter tuning tasks achieving a solve rate of $\geq$ 90\% for $V_{\mathrm{out}}$, $\Delta I_{L}$, $\Delta V_{\mathrm{out}}$, $t_{\mathrm{start}}$ tasks and approximately 80\% for $f_{\mathrm{sw}}$ tasks. For topology adaption tasks on easy and medium-level circuits, its performance is in the range 80\% - 90\%. In contrast, the solve rate declines for topology adaption tasks on the complex LTC7802 circuits, where the performance drops to approximately 50\%.

\paragraph{Refinement of analytical solutions via simulation:}
Figure \ref{fig:question_categories} shows that most output voltage ripple $\Delta V_{\mathrm{out}}$ tuning tasks fail, if no SPICE simulation feedback is integrated into the LLM-based workflow. This observation also applies to tasks related to adjusting the startup time in LTC7802 circuits.\\
One reason for this are the provided datasheet formulas for the voltage ripple \eqref{eq:deltaV} and the startup time, which are only approximations. In many cases these approximations are not precise enough and require iterative refinement through simulation.
\paragraph{Reasoning chain length:}
For the LLM-based workflows without simulation feedback, Figure \ref{fig:question_categories} (orange bars) shows a trend where the solve rate declines from $V_{\mathrm{out}}$ tasks, to $\Delta I_{L}$ tasks, and further to $\Delta V_{\mathrm{out}}$ design tasks.\\
An explanation for this trend are the different "lengths of the reasoning chain" required to solve these tasks. As explained in Section \ref{subsection:benchmarking}, determining the output voltage $V_{\mathrm{out}}$ requires only one equation \eqref{eq:vout}, while calculating the current ripple $\Delta I_{L}$ requires two equations \eqref{eq:vout}, \eqref{eq:deltaI}, while the output voltage ripple $\Delta V_{\mathrm{out}}$ requires solving and correctly combining three equations \eqref{eq:vout} - \eqref{eq:deltaV}.

\begin{figure}[h]
  \centering
  \includegraphics[width=0.6\linewidth]{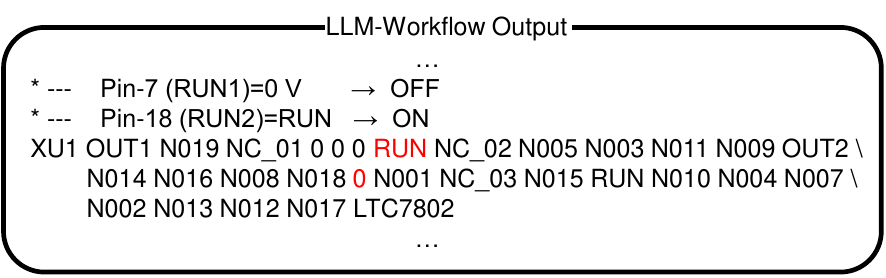}
  \caption{A segment from the LLM output on the task of deactivating the first controller channel and activating the second. In a comment (lines starting with * ), the LLM states correctly that to deactivate the first channel, pin 7 should be connected to ground, and to activate the second channel, pin 18 should be connected to the RUN net. However, the LLM fails to translate this from the natural language comment into proper SPICE netlist syntax}
  \label{fig:example_output}
\end{figure}

\paragraph{Topology adaption tasks:}
In topology adaption, Figure \ref{fig:question_categories} shows that the best workflow (o3 + RAG + SPICE) achieves a success rate of over 80\% on easy and medium circuits, but its performance drops significantly to approximately 50\% on the more complex LTC7802 circuits.\\
Therefore, topology adaption design tasks represent a main limitation for the investigated LLM-based workflows.
We attribute this poor performance primarily to the SPICE netlist syntax, which is unsuitable for LLMs for two reasons. First, the netlist encodes the circuit topology in a tabular format rather than in a language-like structure, see Figure \ref{fig:colored_prompt}. Second, the semantic description of component functionality is weak. For example, the netlist format misses designators for component pins (e.g., "anode," "cathode," "drain"). This causes several errors such as incorrect diode orientation or faulty MOSFET pin connections and controller pin configurations. Figure \ref{fig:example_output} illustrates this problem: the LLM-based workflow correctly describes a required change in a natural-language comment but then fails to translate that instruction into the proper netlist syntax. 

\subsubsection{Computational Cost and Runtime}
\begin{figure}[h!]
    \centering
    \begin{minipage}{0.45\textwidth}
        \centering
        \includegraphics[width=\linewidth]{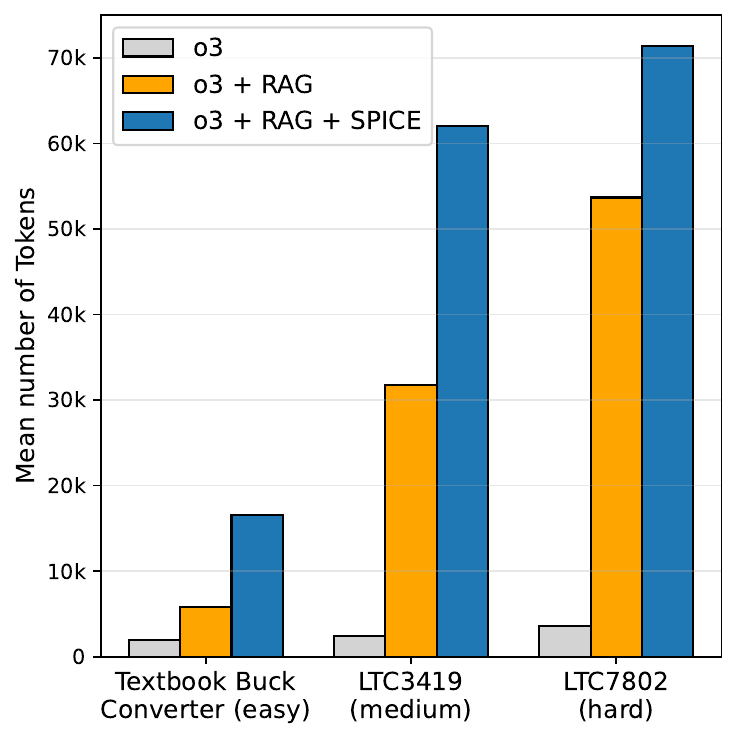}
        \caption{Token usage of different LLM workflows across circuit types with varying complexity}
        \label{fig:toke_usage}
    \end{minipage}%
    \hspace{0.02\linewidth}
    \begin{minipage}{0.45\textwidth}
        \centering
        \includegraphics[width=\linewidth]{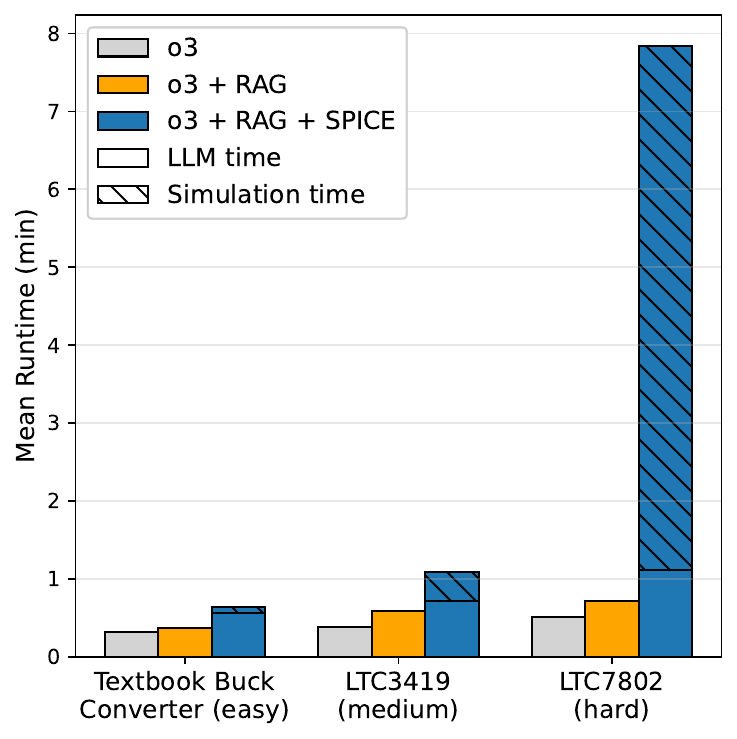}
        \caption{Runtime of LLM and Simulation of different LLM workflows across circuit types with varying complexity}
        \label{fig:runtime} 
    \end{minipage}
\end{figure}

Figure \ref{fig:toke_usage} summarizes the average LLM token usage for the benchmarking tasks in this first experiment, broken down by circuit type and workflow. Token usage increases with circuit complexity. The use of RAG and SPICE simulation tools both significantly increase token count, especially over multiple iterations. This is because RAG adds retrieved document chunks to the input prompt, while each SPICE iteration requires creating a new netlist and reviewing past changes.
Figure \ref{fig:runtime} shows the required runtime of the LLM-based workflows. Similar to token usage, the runtime increases with circuit complexity, the use of RAG, and especially over multiple simulation iterations. For the more complex LTC7802 circuits, the SPICE simulation time becomes the dominant factor, accounting for the majority of the total runtime.

\subsection{Experiment 2: Catalogue Component Selection for Efficiency Optimization}
\label{subsection:real_components}
The first experiment demonstrates that the best workflow (o3 + RAG + SPICE) solves most parameter tuning tasks using idealized lumped-element components. In this second experiment, the goal is to select the best combination of real components to optimize the efficiency of an LTC7802 circuit in light load operation. Instead of using simplified lumped-elements as component models, manufacturer-provided SPICE simulation models \cite{simsurf, kemet} are used. In contrast to previous parameters, such as ripple or output voltage, there is no analytical equation for efficiency estimation based on the given component characteristics. The efficiency is influenced by multiple factors, such as $I^{2}R$ losses ($ESR$, $DCR$, $R_{\mathrm{DS(on)}}$), MOSFET gate capacities, switching frequency, and the operating point of the controller. The search space consists of a discrete set of 16 inductors and 6 capacitors, with the option of placing two output capacitors in parallel. The components are presented to the LLM with relevant context information, as shown in Table \ref{tab:components}. To compare the LLM-based results against a reference, we calculated the efficiency of all possible component combinations via brute force. 

\begin{table}[h]
\caption{Example components, which the LLM can choose, with context information}

\footnotesize
\begin{tabularx}{\textwidth}{l r l r r r r }
\toprule
Part Number & Inductance ($\mu$H) & Size Code & Rated Current (mA) & Sat. Current (mA) & DCR ($\Omega$) \\
\midrule
LQH32PZ2R2NN0   & 2.20 & 3225/1210   & 1600 & 1550.0  & 0.0760  \\
LQH3NPH2R2MME   & 2.20 & 3030/1212   & 2100 & 1800.0  & 0.0650  \\
LQH3NPZ1R0MME   & 1.00 & 3030/1212   & 3000 & 2350.0  & 0.0250  \\
1274AS-H-1R5N   & 1.50 & 100100/3939 & 8900 & 15300.0 & 0.0064  \\
\vdots & \vdots & \vdots  & \vdots & \vdots & \vdots &  \\
\bottomrule
\end{tabularx}
\label{tab:components}
\end{table}

\begin{figure}[h]
  \centering
  \includegraphics[width=0.65\linewidth]{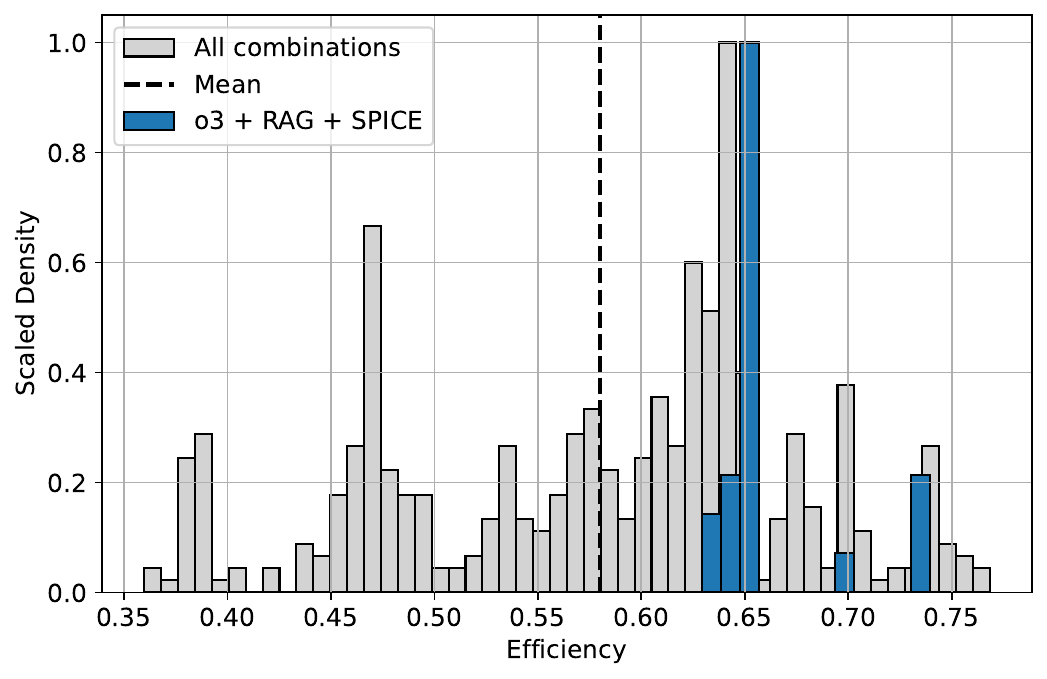}
  \caption{Efficiency of all possible circuits and of the component combinations selected by the LLM-based workflow }
  \label{fig:opt_efficiency}
\end{figure}

Figure \ref{fig:opt_efficiency} shows a histogram of the resulting power conversion efficiencies in 20 independent and identical experiment runs of the LLM-based workflow against the brute-force results. We observe that the component selections suggested by the LLM-based workflow have an efficiency of around 65\%. The efficiency of all possible combinations ranges roughly from 35\% to 75\%. The results suggested by the LLM-based workflow are better than the average solution. Yet, the workflow does not identify the best possible component combination, which could be expected from conventional optimization approaches. 
While the LLM-based workflow is capable to apply general design rules, such as minimizing $ESR$ and $DCR$, it performs an insufficient number of simulation iterations to explore the design space, which prevents it from discovering the optimal solution for these specific conditions in light load operation mode.
The output of the LLM-based workflow can act as reference point for further improvements, as for example in \cite{ledro}, where an LLM is used for design space reduction of optimization techniques.

\section{Discussion}
Our evaluation shows that LLM-based workflows have strong potential to assist in the design process of SMPS circuits. This suggests that the breakthroughs of LLMs in domains such as software development are, to a certain extend, transferable to the hardware development of circuits for PCB applications.\\
Two main factors that are crucial for good performance are simulation feedback and LLM-reasoning. The benchmark experiments demonstrate that the LLMs can effectively integrate SPICE simulation feedback to improve its responses by correcting false assumptions about the circuit or related equation parameters. The SPICE feedback is also used to refine results derived from approximate analytical formulas and to verify "long reasoning chains". This enables an iterative improvement, where the solve rate increases significantly within a small number of iterations. Internal reasoning substantially enhances the performance of LLM-based workflows because it handles required substeps, divides larger tasks in subtasks, and helps the LLM to consider a broader range of possible solution paths.\\
The benchmarking experiment shows that the best LLM workflow - o3 + RAG + SPICE - solves the majority of given parameter tuning tasks for circuits with idealized lumped-element components. The main remaining limitation is handling topology adaption for larger, more complex circuits. Our analysis indicates that one primary reason for this limitation is the syntax of the text-based circuit representations. The SPICE netlist format is disadvantageous for LLMs  because its tabular structure is not language-like, and it has a weak semantic description of component functionality, for example the format lacks features such as pin designators. As a result the LLM often fails to transfer a correct solution, which it is able to describe in natural language, into the netlist format.\\ 
Therefore, our results motivate future research to develop improved text-based circuit representations that are better suited for LLM-based workflows as an alternative to SPICE netlists. First proposals in this direction have been made by \cite{artisan, analogcoder, jitx}. Future research could also investigate the combination of LLMs with conventional optimization algorithms to achieve an optimal selection of catalog components.

\section{Conclusion}
This paper evaluates the capabilities of LLM-based workflows to assist in the development of SMPS circuits for PCB applications. Multiple workflows are presented that combine reasoning LLMs, retrieval-augmented generation (RAG), and a custom toolkit enabling an LLM to interact with SPICE simulations to estimate the impact of circuit modifications. To evaluate the LLM-based workflows, two experiments are conducted. The first experiment focuses on finding idealized components using a broad benchmark of parameter tuning and topology adaption tasks related to SMPS circuits. The goal of the second experiment is to select an optimal set of catalogue components for circuit optimization.
Our analysis shows the following key insights: First, internal reasoning capabilities of LLMs are essential to handle the multi-step design tasks. Second, SPICE simulation feedback can effectively be integrated into the workflow and provides a significant performance improvement. The combination of reasoning LLM, RAG, and SPICE feedback enables the workflow to solve a large majority of the provided parameter tuning tasks with idealized components.
Remaining challenges include topology adaption in larger circuits and design space exploration for circuit optimization.
To address these limitations, future research could focus on developing improved LLM-compatible circuit representations and integrating conventional optimization methods to enhance design space exploration.

\bibliographystyle{acm}
\bibliography{sample}

\end{document}